\documentstyle[psfig,twocolumn,prl,aps,amssymb]{revtex}

\begin{document}
\wideabs{
\draft

\title{NMR Determination
of 2D Electron Spin Polarization at
 \mathversion{bold} $\nu = 1/2$}
\author{S. Melinte$^{1,}$\cite{byline}, N. Freytag$^2$, M. Horvati\'{c}$^2$,
C. Berthier$^{2,3}$, L. P. L\'evy$^{2,4}$, V. Bayot$^5$, and M. Shayegan$^6$}
\address{$^1$Unit\'e PCPM, Universit\'e Catholique de Louvain,
B-1348 Louvain-la-Neuve, Belgium\\
$^2$Grenoble High Magnetic Field Laboratory, MPI-FKF and CNRS,
B.P.166, F-38042 Grenoble Cedex 9, France\\
$^3$Laboratoire de Spectrom\'{e}trie Physique, Universit\'e J. Fourier, BP 87,
F-38402 St. Martin d'H\`{e}res, France\\
$^4$Institut Universitaire de France et Universit\'e J. Fourier, BP 41,
F-38402 St. Martin d'H\`{e}res, France\\
$^5$Unit\'e DICE, Universit\'e Catholique de Louvain,
B-1348 Louvain-la-Neuve, Belgium\\
$^6$Department of Electrical Engineering, Princeton University,
Princeton N.J. 08544}

\date{\today}

\maketitle

\begin{abstract}
Using a `standard' NMR spin-echo technique
we determined the spin polarization $\cal P$ of 
two-dimensional electrons, confined to GaAs quantum wells,
from the hyperfine shift of Ga nuclei in the wells.
Concentrating on the temperature 
($0.05\lesssim T\lesssim 10~{\rm K}$)
and magnetic field ($7 \lesssim B\lesssim 17~{\rm T}$)
dependencies of ${\cal P}$ at Landau level filling factor $\nu =1/2$,
we find that the results are described well by a simple model of 
non-interacting composite fermions, 
although some inconsistencies remain when the two-dimensional electron system is 
tilted in the magnetic field.

\end{abstract}

}



The fractional quantum Hall effect (FQHE), observed in
low-disorder two-dimensional electron systems (2DESs)
at low temperature $T$ and high magnetic field $B$,
is one of the most fascinating problems involving
strongly correlated fermions.
Recently, considerable attention has been focused
on the FQHE ground states near half-integer Landau level (LL) 
filling factors ($\nu=1/2, 3/2$, \ldots), where a large body of 
experimental and theoretical results can
be cast in a surprisingly simple picture of {\em non-interacting}
composite fermions
(CFs) with the same charge and spin as electrons~\cite{CF,Jain,HLR}.
The simplest realization of a CF is at $\nu=1/2$
where electrons bind two flux quanta of a fictitious 
Chern-Simons gauge field~\cite{HLR}. 

An important issue in the physics of CFs is the spin polarization of 
the 2DES at half-integer fillings.
Before the remarkable success of the CF model,
theoretical~\cite{SpinReversalTh},
and experimental~\cite{SpinReversalExp} results, pointed
to the possibility of FQHE states
with reversed spins at various fillings ($\nu = 2/3, 2/5$, \ldots).
In an effort to understand the spin configurations of 
these states within the CF picture and particularly, the
spin polarization of the 2DES close to $\nu = 1/2$, Park and Jain~\cite{Park}
introduced a new parameter,
the CF {\em polarization mass} $m_{p}^{*}$, which is proportional to 
the ratio of the cyclotron and Coulomb energies.
These authors obtained an estimate for $m_{p}^{*}$ at $\nu = 1/2$: 
\begin{equation}
m_{p}^{*}/m_{e} \cong  0.60 \sqrt{B_{\perp}},   
\label{eq:mass}
\end{equation}
where $B_{\perp}$ (in tesla) is the component of $B$ perpendicular to the
2DES plane~\cite{NoteI,NoteII}.
The parameter $m_{p}^{*}$, combined with
a parabolic dispersion law for CFs at $\nu =1/2$, uniquely 
determines the spin polarization ${\cal P}$ at any given $T$ and $B$.
Here we report direct measurements of the 2DES spin polarization
as a function of $T$ and $B$, and use our data to critically test the 
applicability of the non-interacting CF model and $m_{p}^{*}$.
We find that the data are in excellent agreement with 
predictions of Ref.~\cite{Park}, except when the 2DES is tilted in $B$.

Nuclear magnetic resonance (NMR) is a sensitive technique for the 
experimental determination of the spin polarization of 
two-dimensional (2D) electrons~\cite{OPNMR1,OPNMR2}.
Prior to this work, however, only the use
of optically pumped NMR (OPNMR) was reported~\cite{OPNMR1,OPNMR2}.
The reason is that the number
of active nuclei in a typical 2D system is usually too small
to generate a useful signal for 'standard' NMR techniques.
One way to increase the signal is by optical pumping:
polarized electrons are excited in the conduction band by
illuminating the sample with circularly
polarized light and the strong hyperfine coupling ensures the
transfer of this polarization to the nuclei.
However, this also implies that in OPNMR experiments
the electronic system is observed while nuclei are strongly polarized,
well beyond their small equilibrium value.
We demonstrate here that the {\it standard} pulsed NMR
technique, applied to the Ga nuclei
in ${\rm GaAs/AlGaAs}$ multiple-quantum well heterostructures does indeed
provide measurable signal for $T\lesssim 10$~K. 
The NMR signal was observed on samples consisting of 200 quantum wells (QWs)
using a state-of-the-art laboratory-built pulsed NMR spectrometer. 
The method we employ
avoids any eventual perturbation of the system by optical pumping
and the experimental setup is greatly simplified,
making possible, e.g., the use of a ${\rm ^{3}He/^{4}He}$
dilution refrigerator.


Two heterostructures, M242 and M280, each composed of
one hundred GaAs QWs, separated
by AlGaAs barriers which are Si-doped near their centers, were used 
in this study~\cite{MQW}.
Sample M242 (M280) has $250 \rm \ \AA$ ($300 \rm \ \AA$) wide QWs,
$1850{\rm \ \AA }$ ($2500{\rm \ \AA }$) thick $\rm Al_{0.3}Ga_{0.7}As$ 
($\rm Al_{0.1}Ga_{0.9}As$) barriers,
and density $n=1.4\times 10^{11}{\rm \ cm^{-2}}$
($8.5\times 10^{10}{\rm \ cm^{-2}}$).
From each heterostructure, we cut two 
$\approx 26 \rm \ mm^{2}$ pieces 
and placed the two pieces together into the radio-frequency
coil, so that experiments were done on effectively  200 QWs.
For $T\gtrsim 1.5$~K, the NMR signal was recorded as a
function of both $B$ and tilt angle $\theta$ between
$B$ and the normal to the plane of the 2DES.
For very low-$T$ measurements, the radio-frequency coil
was mounted into the mixing chamber of a ${\rm ^{3}He/^{4}He}$
dilution refrigerator, and measurements were
performed as a function of $T$ at fixed $\theta$ and $B$.


To distinguish between the contributions of Ga nuclei
in QWs and barriers and to
eliminate the signal from the substrate,
we exploited the difference in their
nuclear spin-lattice relaxation rates ($1/T_{1}$)~\cite{OPNMR1}.
The NMR pulse sequence is described in
Fig.~\ref{fig1}a:  the nuclear magnetization was
first set to zero by a comb of $\pi /2$ pulses.
After the magnetization has recovered during time $t_{{\rm R}}$,
its value was measured by a spin-echo sequence
($\pi /2-\tau -\pi-\tau- \rm{echo}$)~\cite{Abragam}.
Spectra were obtained by Fourier transforming the echo
(Figs.~\ref{fig1}b and ~\ref{fig1}c), and the hyperfine shift
$K_{S}$ of Ga nuclei in QWs is here defined 
as the frequency shift of the NMR line attributed to the 
QWs with respect to the barriers' line.
This resonance shift is caused by the hyperfine interaction between
nuclei and 2D electrons
(dominated by the Fermi contact term)~\cite{OPNMR1,OPNMR2,Abragam}.
Note also that our QWs' NMR line is split by a small and
well defined quadrupole coupling which is clearly resolved at high
temperatures (Fig.~\ref{fig1}b), confirming the high homogeneity
of the 2DES.

Before focusing on the spin polarization at $\nu =1/2$
we first discuss the results at $\nu =1/3$.
Previous OPNMR
experiments~\cite{OPNMR2} revealed a completely spin polarized FQHE 
ground state at $\nu =1/3$.
The low $T$ (${\cal P}=1$)
limit of OPNMR $K_{S}$ data at $\nu =1/3$,
measured for several samples,
was successfully used to determine the
intrinsic hyperfine shift of Ga nuclei in the center of each QW 
[$K_{S\rm {int}}=K_{S}+1.1 \times(1-\exp(-K_{S}/2.0))$]
and to establish the relationship $K_{S\rm {int}}=A_{c}{\cal P}n/w$,
which defines the hyperfine coupling
$A_{c}=(4.5\pm 0.2)\times 10^{-13}{\rm \ cm}^{3}/{\rm s}$ ($w$ is the 
QW width)~\cite{OPNMR2}.
Applied to our samples, these expressions yield the
reference `full polarization' values $K_{S\rm {int}}^{{\cal P}=1}\approx 12.7$~kHz
for M280 and $K_{S\rm {int}}^{{\cal P}=1}\approx 25.2$~kHz for M242~\cite{FullP}.
Figure~\ref{fig2} shows that the
latter value is consistent with our very low $T$ data.
The $T$-dependence of
these data is fitted to $K_{S\rm {int}}^{sat}(\nu \cong 1/3)\tanh (\Delta
_{1/3}/4T)$, yielding $K_{S\rm {int}}^{sat}(\nu \cong 1/3)=21\pm 2.5\ {\rm kHz}$ 
and $\Delta_{1/3}=1.7\Delta _{Z}$, in
agreement with the OPNMR result 
$\Delta_{1/3}=1.82\Delta _{Z}$~\cite{OPNMR2} and theoretical
estimates $\Delta_{1/3}\approx 2\Delta _{Z}$~\cite{MacDonald}.
Here $\Delta _{Z}=|g|\mu_{B}B$ is the Zeeman energy, $\mu_{B}$ is 
the Bohr magneton, and $g=-0.44$ is the electron $g$-factor in bulk GaAs.

In Fig.~\ref{fig3} we present the  $T$-dependence of 
the 2DES spin polarization at $\nu =1/2$ at different $B$ for our two 
samples.
The right axes give the measured $K_{S\rm {int}}$, 
while the deduced spin polarization, defined as 
${\cal P}(T) =K_{S\rm {int}}(T)/K_{S\rm {int}}^{{\cal P}=1}$,
is indicated on the left axes. 
Concentrating on the $\theta =0^{\circ}$ data (filled circles),
we note that $K_{S\rm {int}}$
for the high density 2DES (M242)
at $B=11.4$~T reaches the full polarization value as $T\rightarrow 0$, implying
that the ground state of the 2DES is fully spin polarized at 
$\nu = 1/2$.
On the other hand,
M280 data at $B=7.1$~T reveal that the low-$T$  $K_{S\rm {int}}$ saturates at 
$9.5 \pm 1$~kHz,
below the expected $K_{S\rm {int}}^{{\cal P}=1}\approx 12.7$~kHz for this sample.  
The ground state in M280
therefore appears to be only {\em partially} spin polarized at $\theta =0^{\circ}$.  

In the remainder of the paper we will discuss how these conclusions,
as well as the $T$-dependencies reported in Fig.~\ref{fig3},
compare to the non-interacting CF model of Ref.~\cite{Park}. 
However, without referring to any model, we can already infer useful
information from the data presented so far by considering the
Zeeman energy normalized to the Coulomb energy ($\Delta_{C}$) for the
two samples. Here $\Delta_{C}= e^2/ \epsilon l_{B}$,
$\epsilon \approx 13$ is the static dielectric constant of GaAs, 
and $l_{B}=\sqrt{\hbar/e B_{\perp}}$ is the magnetic length.
The $\Delta_{Z}/\Delta_{C}$ ratio
is 0.019 for M242 at $B=11.4$~T and 0.016 for M280 at $B=7.1$~T,
implying that the 2DES becomes fully
spin polarized for $\Delta_{Z}/\Delta_{C}$
above a critical value which lies between 0.016 and 0.019.  
This conclusion is consistent with magneto-optics data~\cite{Kukushkin} which
yielded a critical value of 0.018 for the full spin polarization at $\nu =1/2$.

We now attempt to understand the $T$-dependence of ${\cal P}$ based 
on a simple model of non-interacting CFs. 
We assume that, consistent with
previous work~\cite{CF}, CFs have a $g$-factor
roughly the same as electrons and consider parabolic bands occupied by 
$n$ CFs with mass $m_{p}^{*}$.  
Hence, the density of states $D_{\pm }(E)$ (for spin-up
and spin-down CFs) is $D_{\pm }(E)=D\vartheta (E\pm \Delta _{Z}/2)$,
where $D=m_{p}^{*}/(2\pi \hbar ^{2})$\ and $\vartheta $\ is the step function.
Making use of Fermi-Dirac distribution we find
\begin{equation}
{\cal P}=\frac{D}{n}\left[ \Delta _{Z}-2k_{B}T\ \cdot \ \tanh^{-1}\left( {1+\frac{%
\exp (\frac{n}{Dk_{B}T})}{\sinh ^{2}(\frac{\Delta _{Z}}{2k_{B}T})}}\right)
^{-\frac{1}{2}}\right].   \label{eq:P}
\end{equation}
Depending on the strength of the magnetic field, this model
predicts either partially or completely polarized Fermi sea of CFs in the
$T\rightarrow 0$ limit, i.e., ${\cal P}(T=0)=\min \{D\Delta_Z/n,1\}$.
Note that according to Eq.~\ref{eq:P}, ${\cal P}$ at a given $T$ and $B$
depends {\it only} on $m_{p}^{*}$.
Taking $m_{p}^{*}$ as a fitting parameter, in Fig.~\ref{fig3}
we show the best fits of Eq.~\ref{eq:P} to
the $\theta =0^{\circ}$ data by solid curves.
These curves indeed provide  a reasonable description of the data. 
Moreover, the deduced $m_{p}^{*}$ values ($2.2m_{e}$ at $B=11.4$~T 
and $1.7m_{e}$ at $B=7.1$~T)
are found to be in excellent agreement with the
polarization mass predicted by Eq.~\ref{eq:mass}: 
$m_{p}^{*}/m_{e}=2.0$ and 1.6 at $B=11.4$~T and $B=7.1$~T, respectively.
This agreement is quite remarkable as it implies that $m_{p}^{*}$ given by
Eq.~\ref{eq:mass}
together with the simple model leading to Eq.~\ref{eq:P} give a very good account of
the $\theta =0^{\circ}$ data without any adjustable parameters.

Next, we present our study of ${\cal P}$ in tilted magnetic fields
(Figs.~\ref{fig3} and ~\ref{fig4}).
Unfilled circles in Fig.~\ref{fig3} show $\nu =1/2$ data
at $\theta = 40^{\circ}$ (M242) and
$\theta = 61^{\circ}$ (M280), taken as a function of $T$ at $B=14.8$~T.
The data for both samples have a qualitatively similar behavior: the 
polarization at $\theta \neq 0^{\circ}$ is larger than at $\theta =0^{\circ}$ for 
low and intermediate $T$ while at highest 
$T$ the measured polarization falls below the $\theta =0^{\circ}$ values.
In Fig.~\ref{fig4} we present data, taken at $T\approx 1.5$~K, showing the 
dependence of ${\cal P}$ on $B$ for both M242 and M280. 
Here the spin polarization 
exhibits a monotonic increase with $B$.

To compare these data with the predictions of the non-interacting CF model,
we show in Figs.~\ref{fig3} and ~\ref{fig4} plots of ${\cal P}(T,B)$
according to Eq.~\ref{eq:P}.  
Note that here there are no adjustable parameters as we used the
$m_{p}^{*}$ values obtained from the fits to the $\theta =0^{\circ}$ data.  
The only parameter in Eq.~\ref{eq:P} that depends on the total 
magnetic field  is $\Delta_{Z}$.  
The results (dotted curves in Fig.~\ref{fig3})
qualitatively agree with the data in the low and intermediate $T$ range where
both the measured and calculated ${\cal P}$ lie above the $\theta =0^{\circ}$ data.  
However, the calculated ${\cal P}$ overestimates the measured ${\cal P}$ in the
entire $T$ range~\cite{Fits}.
The data of Fig.~\ref{fig4}, on the other hand, at first sight 
appear to be in reasonable agreement with Eq.~\ref{eq:P}. 
But it is likely that this agreement is fortuitous, 
as it occurs at a particular intermediate temperature.
Note also that in Fig.~\ref{fig4} the difference between the calculated
curves and the data becomes larger at higher $B$ (larger tilt angles).
These observations suggest that the simple CF model leading
to Eqs.~\ref{eq:mass} and \ref{eq:P} are not directly applicable at large $\theta$.  
The deformation of the 2DES
wavefunction at large $\theta$
may be partly responsible for this discrepancy, although
we cannot rule out other possibilities.


In conclusion, we demonstrated the feasibility of the `standard' NMR
experiments to investigate the spin polarization of the 2DES in the 
quantum limit.
Our results provide experimental support for the simplest and most straightforward 
model of ${\cal P}(T,B)$ at $\nu =1/2$.
This model, based on the assumption of a parabolic dispersion law 
of CFs with a polarization effective mass $m_{p}^{*}$, describes well 
the experimental data when the 2DES is subjected only to a 
perpendicular magnetic field.


We thank P. van der Linden for help with the experiment and D. Haldane
for the enlightening discussion.
This work has been partially supported by NATO grant CRG 950328,
by the NSF MRSEC grant DMR-9809483,
and by the "Communaut\'{e} Fran\c{c}aise de Belgique'
(directly and via `PAI' program).
M. Shayegan acknowledges support by the Alexander von Humboldt Foundation.


{\em Note added.-} As this manuscript was being completed we became aware of
similar work~\cite{OPNMR3} 
investigating the spin polarization at $\nu=1/2$ by OPNMR.



\begin{figure}
\caption{(a) Pulse sequence used to detect the NMR spectra. 
${\rm ^{71}Ga}$ NMR spectra taken on M242 at 
(b) $f_{0}=73.915{\rm \ MHz}$ with $t_{\rm R}$ = 256\ s (top) and 2\ s (bottom)
and at (c) $f_{0}=192.052{\rm \ MHz}$ with $t_{\rm R}$ = 128\ s (top) and 
32\ s (bottom).
For short recovery times (lower spectra) the contribution is
essentially from nuclei in QWs, while for longer times (upper spectra)
barriers' signal becomes stronger than the one from QWs.}
\label{fig1}
\end{figure}

\begin{figure}
\caption{${\rm ^{71}Ga}$ intrinsic hyperfine shift ($K_{S\rm {int}}$) vs $T$ 
for M242 at $\protect\theta =0^{\circ}$ and
$B=17\ {\rm T}$ ($\nu =0.335$).
The solid curve is a fit to the data (see text).}
\label{fig2}
\end{figure}

\begin{figure}
\caption{${\cal P}$ (left axes) and  $K_{S\rm {int}}$
(right axes) vs $T$ at
$\protect\nu =1/2$ for M242 (top panel) and M280 (bottom panel).
The filled circles represent $\protect\theta =0^{\circ}$ data
and the unfilled circles represent $\protect\theta \neq 0^{\circ}$ data.
In both panels the solid  
curves represent best fits of Eq.~\ref{eq:P} to the $\protect\theta =0^{\circ}$ 
data; these fits give $m_{p}^{*}/m_{e}=2.2\pm 0.2$ for M242 and
$m_{p}^{*}/m_{e}=1.7\pm 0.2$ for M280.
The dotted curves represent predictions of Eq.~\ref{eq:P} for $B=14.8\ {\rm T}$
and using $m_{p}^{* }=2.2m_{e}$ (M242) and $m_{p}^{* }=1.7m_{e}$ (M280).}
\label{fig3}
\end{figure}

\begin{figure}
\caption{${\cal P}$ vs $B$
at $\nu =1/2$ and $T \approx 1.5$~K for M242 ($\circ $) and M280 ($\bullet $).
The dotted and solid curves represent
predictions of Eq.~\ref{eq:P}. The solid curve was computed
with $m_{p}^{* }=2.2m_{e}$ (M242), and the dotted one 
with $m_{p}^{* }=1.7m_{e}$ (M280).}
\label{fig4}
\end{figure}

%
%


\begin{references}
\bibitem{NoteI}The polarization mass $m_{p}^{*}$ in Eq.~\ref{eq:mass}
is essentially a CF density of states effective mass. 
It is much larger than the CF 'activation mass'
which determines the activation energies of the FQHE states in the CF 
picture~\cite{Park}.
\bibitem{FullP} We have an uncertainty of $\pm 10$\% in our 
samples' $w$, resulting in a similar uncertainty in our
calculated $K_{S\rm {int}}^{{\cal P}=1}$.
\bibitem{Fits}  We note that Eq.~\ref{eq:P}
does not fit the 14.8~T data of Fig.~\ref{fig3}, even if we take
$m_{p}^{*}$ to be a fitting parameter:
the curves (not shown) fit the data points only up to about 2~K
but fall well above them at higher $T$.
We also tried fitting the data of Fig.~\ref{fig3} using both $m_{p}^{*}$
and the $g$-factor ($\Delta_{Z}$) as fitting parameters. 
The results of two-parameter fits to the $\theta =0^{\circ}$ data are 
$|g|=0.42$ and $m_{p}^{*}/m_{e}=2.4$ for M242,
and $|g|=0.39$ and $m_{p}^{*}/m_{e}=2.0$ for M280.  
For the tilted $B$ data, we obtain a reasonable fit for
M242 with $|g|=0.29$ and $m_{p}^{*}/m_{e}=4.0$ (which is much larger than
the $\theta =0^{\circ}$ value).
But for M280, with $|g|=0.36$ and $m_{p}^{*}/m_{e}=1.2$ (which is much smaller than the
$\theta =0^{\circ}$ value), the fitted curve still deviates substantially from the
experimental data. 
We conclude that Eq.~\ref{eq:P} does not describe the experimental
data taken in tilted $B$.
\end{references}
\end{document}